\documentclass[prd,twocolumn,superscriptaddress,nofootinbib,showpacs]{revtex4}

\usepackage{graphicx,amsmath,amsfonts,amssymb,revsymb,dcolumn,epsfig,bm}

\begin{document}
\title{Energy Conditions and Supernovae Observations}

\author{J. Santos} \email{janilo@dfte.ufrn.br}
\affiliation{Universidade Federal do Rio Grande do Norte \\ Departamento de
F\'{\i}sica \\ C.P. 1641, 59072-970 Natal -- RN, Brasil}

\author{J.S. Alcaniz}\email{alcaniz@on.br} 
\affiliation{Departamento de Astronomia, Observat\'orio Nacional\\
20921-400, Rio de Janeiro -- RJ, Brasil}

\author{M.J. Rebou\c{c}as}\email{reboucas@cbpf.br} 
\affiliation{Centro Brasileiro de Pesquisas F\'{\i}sicas\\
22290-180 Rio de Janeiro -- RJ, Brasil}

\date{\today}
 
\begin{abstract}
In general relativity, the energy conditions are invoked to restrict general energy-momentum tensors $T_{\mu\nu}$ on physical grounds. We show that in the standard Friedmann--Lema\^{\i}tre--Robertson--Walker (FLRW) approach to cosmological modelling where the equation of state of the cosmological fluid is unknown, the energy conditions provide model-independent bounds on the behavior of the distance modulus of cosmic sources as a function of the redshift. We use both the  \emph{gold} and the \emph{legacy} samples of current type Ia supenovae to carry out a model-independent analysis of the energy conditions violation in the context of standard cosmology.
\end{abstract}

\pacs{98.80.Es, 98.80.-k, 98.80.Jk}

\maketitle

\section{Introduction}


The standard approach to cosmological modelling commences with the assumption that our $3$--dimensional space is homogeneous and isotropic at large scales. The most general spacetime metric consistent with the existence of a cosmic time $t$ and the principle of spatial homogeneity and isotropy is the FLRW metric
\begin{equation}
\label{RWmetric} ds^2 = - dt^2 + a^2 (t) \left[\, d \chi^2 +
S_k^2(\chi)\, (d\theta^2 + \sin^2 \theta  d\phi^2) \,\right],
\end{equation}
where the function $S_k(\chi)=(\chi\,$, $\sin\chi$, $\sinh\chi)$ depends on the sign of the constant spatial curvature ($k=0,1,-1$), $a(t)$ is the cosmological scale factor, and we have set the speed of light $c = 1$. A third assumption in this approach is that  the large scale structure of the universe is essentially determined by gravitational interactions, and hence can be described by a  metrical theory of gravitation such as  General Relativity (GR), which we assume in this work.

Under these very general premises, the total density $\rho$ and the total pressure $p$ of the cosmological fluid as a function of scale factor $a$ are given by 
\begin{eqnarray}
\rho & = & \frac{3}{8\pi G}\left[\,\frac{\dot{a}^2}{a^2}
                                        +\frac{k}{a^2} \,\right]\;,
\label{rho-eq} \\
p & = & - \frac{1}{8\pi G}\left[\, 2\,\frac{\ddot{a}}{a} +
\frac{\dot{a}^2}{a^2} + \frac{k}{a^2} \,\right] \;, \label{p-eq}
\end{eqnarray}
where $G$ is the Newton constant. 

Note that  if now one wishes to constrain the physical properties that hold for the matter fields in the Universe it is convenient to impose the so-called \emph{energy conditions} \cite{Hawking-Ellis,Visser,Carroll} that limit the arbitrariness of the energy-momentum tensor $T_{\mu\nu}$ of these fields or, equivalently, of the physical behavior of their associated energy density $\rho$ and pressure $p$. These conditions can be stated in a coordinate-invariant way, in terms of $T_{\mu\nu}$ and vector fields of fixed character (timelike, null and spacelike). The most common energy conditions are\footnote{Although physically well-motivated, over the past years views have changed as to how fundamental some of the specific energy conditions are, and by the end  the 1970's it became clear that not all forms of matter 
sources obey the energy conditions. The cosmological constant $\Lambda$ is perhaps the best known example of matter fields that violates the strong energy condition, but fulfills the weak energy condition ($\rho \geq 0$ and $\rho + p \geq 0$). See, e.g. \cite{Visser-Barcelo,Alametal} for a short historical review.}:
\begin{itemize}
\item[(i)] The null energy condition (NEC). NEC states that
$T_{\mu\nu}n^{\mu}n^{\nu}\geq 0$ for null vectors $n^{\mu} \in T_{s}(M)$, where $M$ is a real 4-dimensional space-time manifold and $T_{s}(M)$ denotes the tangent space to $M$ at a point $s \in M$.

\item[(ii)] The weak energy condition (WEC). WEC states that
$T_{\mu\nu}t^{\mu}t^{\nu}\geq 0$ for any timelike vector $t^{\mu} \in T_{s}(M)$. This
will also imply, by continuity, the NEC.

\item[(iii)] The strong energy condition (SEC). SEC is the assertion
that for any timelike vector $(T_{\mu\nu} - T/2g_{\mu\nu})t^{\mu}t^{\nu}\geq 0$,
where $T$ is the trace of $T_{\mu\nu}$.

\item[(iv)] The dominant energy condition (DEC). DEC requires that
$T_{\mu\nu}t^{\mu}t^{\nu}\geq 0$ for any timelike vector $t^{\mu} \in T_{s}(M)$ and the
additional requirement that $T_{\mu\nu}t^{\mu}$ be a non-spacelike vector.
By continuity this will also hold for any null vector $n^{\mu} \in T_{s}(M)$.
\end{itemize}

In terms of the energy-momentum tensor for a perfect fluid, $T_{\mu\nu} = (\rho+p)\,u_\mu u_\nu + p \,g_{\mu \nu}$,
the  above conditions  (see, e.g., \cite{Hawking-Ellis,Visser,Carroll,Visser-Barcelo,Alametal,carroll1}) reduce to 
\[
\begin{array}{llll}
\mbox{\bf WEC} \ & \Longrightarrow & \rho \geq 0 &
\ \mbox{and} \quad\, \rho + p \geq 0 \;,  \\
\\
\mbox{\bf SEC}   &\Longrightarrow & \rho + 3p \geq 0 &
\ \mbox{and} \quad\, \rho + p \geq 0 \;, \\
\\
\mbox{\bf DEC}   &\Longrightarrow & \rho \geq 0  &
\ \mbox{and} \; -\rho \leq p \leq\rho \;,
\end{array}
\]
where the NEC restriction ($\rho+p \geq 0$) has been incorporated by the WEC.

Thus, from  equations (\ref{rho-eq})--(\ref{p-eq}) one can easily rewrite the energy conditions as a set of dynamical constraints involving the scale factor $a(t)$ and its derivatives for any spatial curvature, i.e.,
\begin{eqnarray}
\mbox{\bf WEC} & \, \Longrightarrow & \quad\, - \frac{\ddot{a}}{a}
+  \frac{\dot{a}^2}{a^2}  + \frac{k}{a^2} \geq 0 \;, \label{wec-eq} 
\\
\mbox{\bf SEC} & \, \Longrightarrow & \qquad \frac{\ddot{a}}{a} \leq 0 \;, 
\label{sec-eq} \\
\mbox{\bf DEC} & \, \Longrightarrow & \; - 2\left[ 
\frac{\dot{a}^2}{a^2}+\frac{k}{a^2} \right] \leq 
\frac{\ddot{a}}{a}  \leq 
\frac{\dot{a}^2}{a^2}+\frac{k}{a^2}  \label{dec-eq}.
\end{eqnarray}
Clearly, in a expanding FLRW universe  the SEC implies that the expansion of the universe is decelerating irrespective of the sign of the spatial curvature. Furthermore,
since most of the ordinary forms of matter obey the DEC they inevitably fulfill the less restrictive WEC. However, violation of the DEC does not necessarily implies violation of WEC or even of the SEC.

In this \emph{Brief Report}, to shed some light on the energy conditions interrelations from 
an observational viewpoint, we use the above 
dynamical formulation of the energy conditions to derive 
model-independent bounds on the luminosity distance $d_L$ of cosmic sources 
in the expanding FLRW flat universe. We then concretely confront  these bounds with current SNe Ia observations, as provided by the \emph{High-z Supernovae Team} (HzST) \cite{Riess2004} and the \emph{Supernova Legacy Survey} (SNLS) collaboration \cite{Legacy2005}.

\section{Distance Modulus Bounds from Energy Conditions}

In practice, distance measurements to distant sources are made in terms of \emph{distance modulus\/}
$\,m - M$, where $m$ is the apparent magnitude of the source and
$M$ its absolute magnitude. As it is well known, the distance modulus is related to
the luminosity distance $d_L$ via\footnote{A related study involving the energy conditions constraints on the lookback time-redshift relation and the epoch of galaxy formation is found in Ref.~\cite{visserS}. For other application involving energy conditions see, e.g.,
Ref.~\cite{Ref10}.}
\begin{equation} \label{dist-mod}
\mu(z) \equiv m(z) - M = 5\,\log_{10} d_L (z) + 25\;,
\end{equation}
where $d_L(z)$ is measured in Mpc. In an expanding FLRW spatially 
flat universe $d_L(z)$ is given by
\begin{equation}  \label{dist-lumin}
d_L(z) = a_0(1+z)\int_a^{a_0}\,\frac{da}{a\,\dot{a}} \;,
\end{equation}
where the subscript $0$ stands for the present-day quantities. In what follows we focus our attention only on the flat
($k=0$) case. 

\begin{figure}[t]
\begin{center}
\includegraphics[width=6.5cm,height=8.0cm,angle=270]{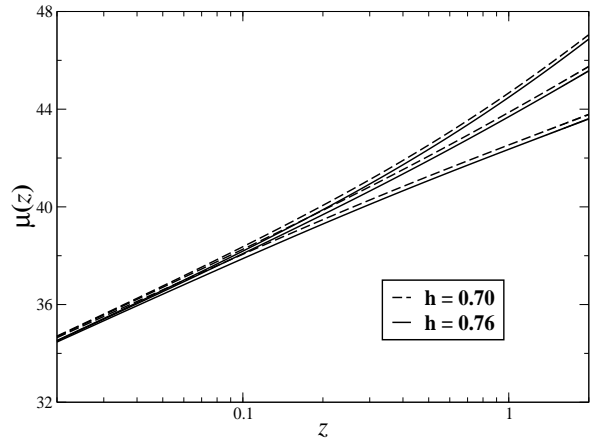}
\caption{\label{Econd-SNLS} Model-independent bounds on the distance modulus $\mu(z)$ as a function of the redshift for two different values of the Hubble parameter, i.e, $h = 0.70$ (dashed lines) and $h = 0.76$ (solid lines). From top to bottom the curves correspond to the WEC, SEC and DEC $\mu(z)$ predictions. Note that, in the \emph{hierarchy} of the predictions for $\mu(z)$, violation of the WEC also means violation of both SEC and DEC.}
\end{center}
\end{figure}

\vspace{0.2cm}

\centerline{\bf{A.  WEC}}

\vspace{0.2cm}

In order to obtain the bounds provided by the WEC on $\mu(z)$ we note that eq.~(\ref{wec-eq}) 
can be written as $d(\dot{a}/a)/\,dt \leq 0$ or, equivalently, 
\begin{equation} \label{wec-ineq}
\mbox{\bf WEC} \quad \, \Rightarrow \quad\, \dot{a} \geq H_0\,a
\quad \forall \quad a < a_0,
 \end{equation}
where  $\dot{a}(a)$ is the velocity of expansion of the universe 
as a function of scale factor, and $H_0=\dot{a}(t_0)/a(t_0)$ is 
the Hubble parameter today.
Now, making use of the inequality~(\ref{wec-ineq}) we integrate~(\ref{dist-lumin})
to obtain the following upper bound for the distance modulus: 
\begin{equation}  \label{WEC-bound}
\mbox{\bf WEC} \quad \Longrightarrow \quad \mu(z) \leq 5\,\log_{10}\,
\left[\,H_0^{-1}\,z(1+z) \,\right]+25,
\end{equation}
where we have used that $a_0/a = 1 + z\,$. Clearly,  if the WEC is obeyed 
then $\mu(z)$ must take values such that Eq.~(\ref{WEC-bound}) holds.  Note also that the condition $\rho>0$ gives no further restrictions on
the distance modulus, and therefore the WEC and NEC bounds on $\mu(z)$
coincide [Eq. (\ref{WEC-bound})].

\begin{figure*}[htb!]
\begin{center}
\includegraphics[width=6.5cm,height=5.8cm,angle=-90]{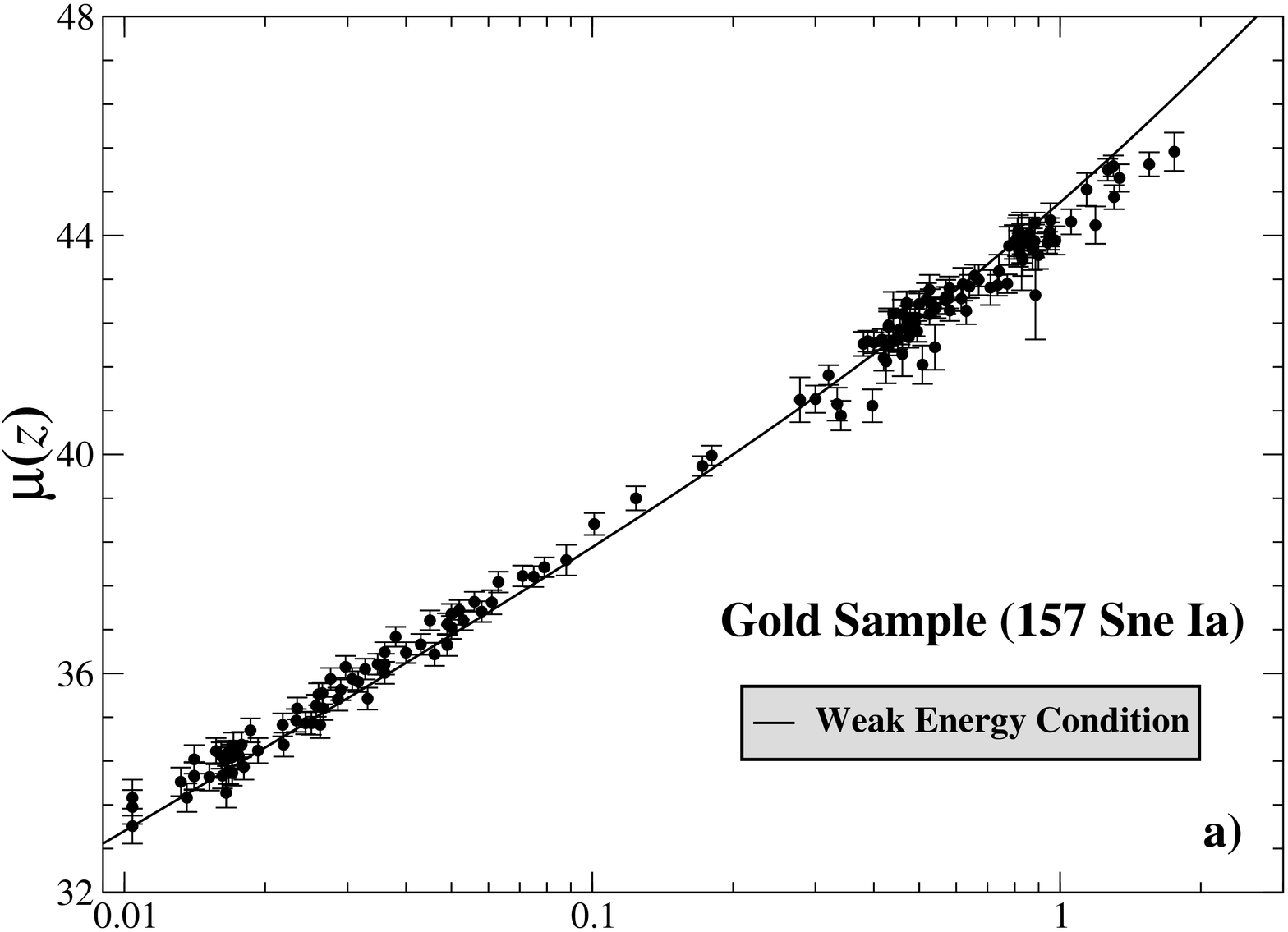}
\includegraphics[width=6.5cm,height=5.8cm,angle=270]{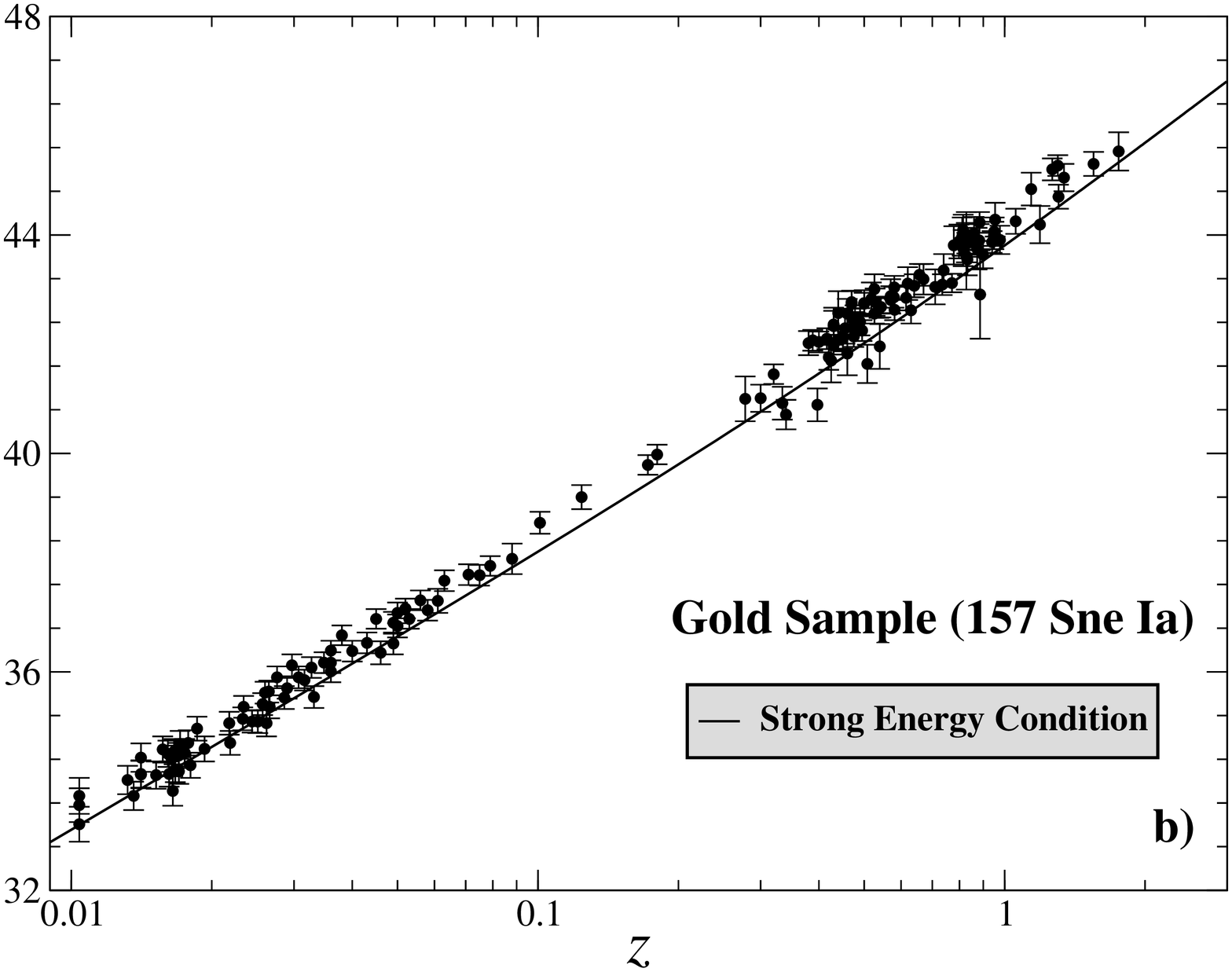}
\includegraphics[width=6.5cm,height=5.8cm,angle=270]{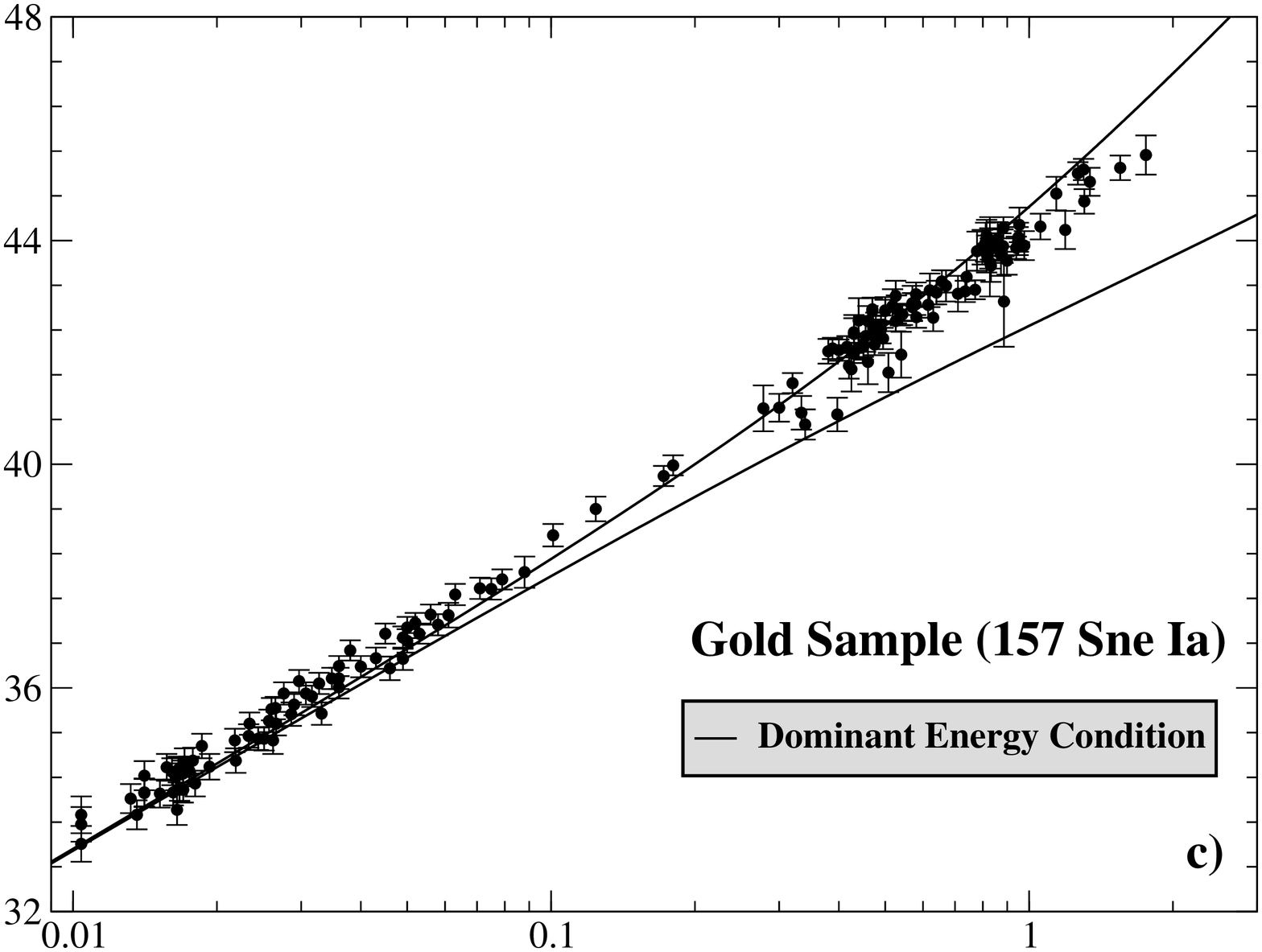}
\includegraphics[width=6.5cm,height=5.8cm,angle=270]{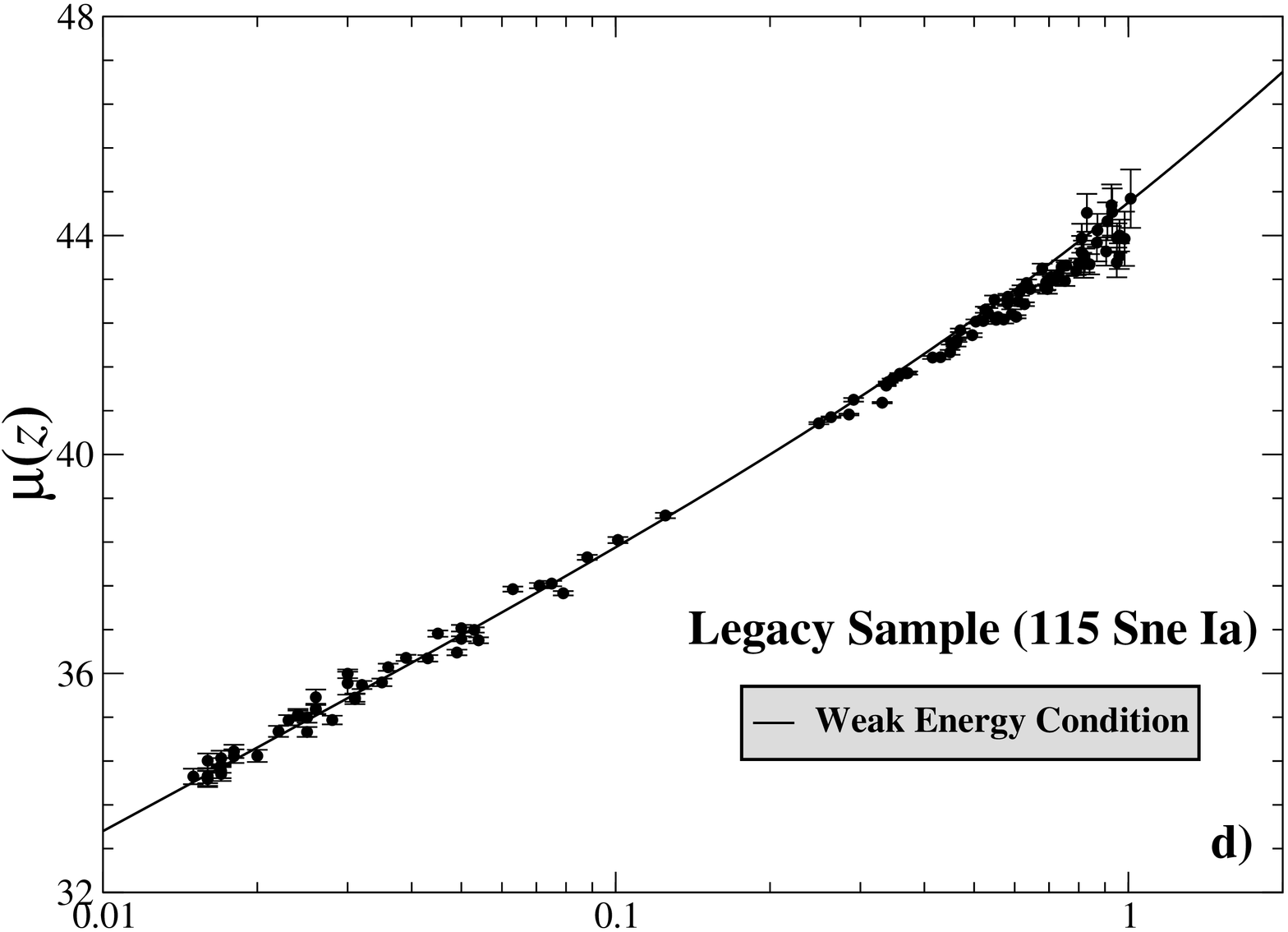}
\includegraphics[width=6.5cm,height=5.8cm,angle=270]{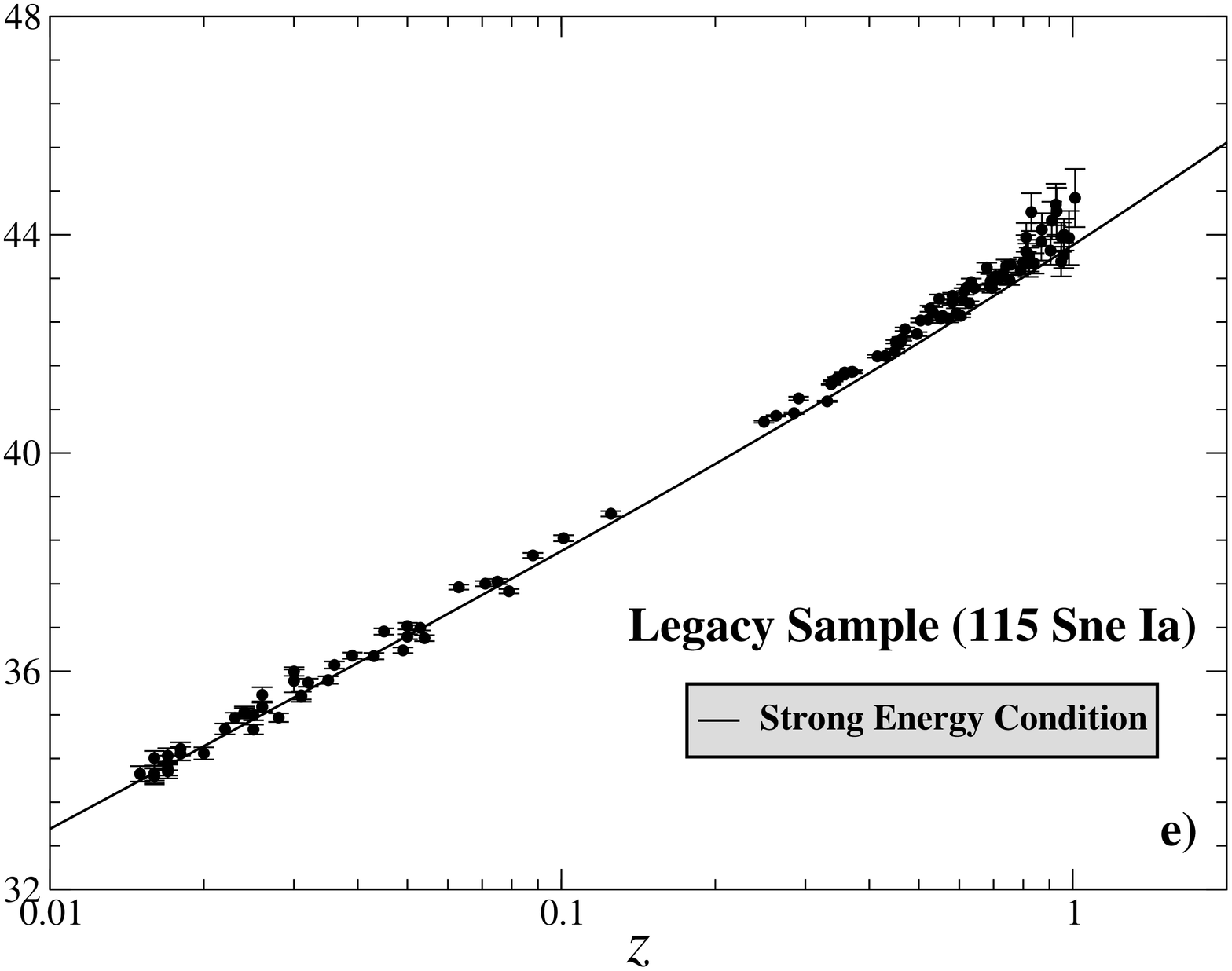}
\includegraphics[width=6.5cm,height=5.8cm,angle=270]{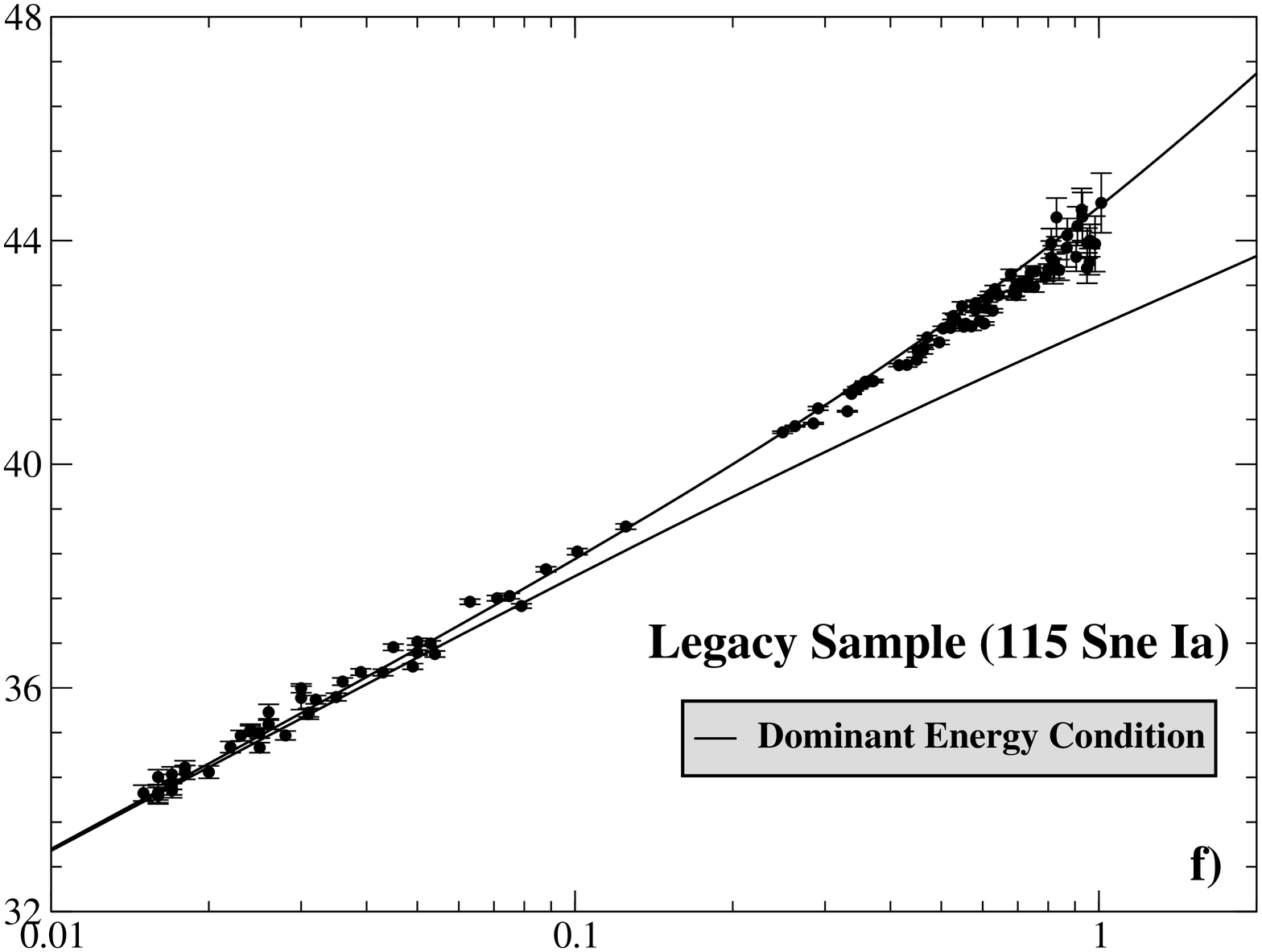}
\caption{\label{Econd-Gold} Energy conditions predictions for $\mu(z)$. The data points appearing in the first row (Panels 2a, 2b and 2c) correspond to the so-called \emph{gold} sample \cite{Riess2004}, whereas in the second row (Panels 2d, 2e and 2f) the data set corresponds to the first year results of the SNLS collaboration \cite{Legacy2005}. As discussed in the text, all the energy conditions seem to have been violated in a recent past of the cosmic evolution.}
\end{center}
\end{figure*}

\vspace{0.2cm}

\centerline{\bf{B.  SEC}}

\vspace{0.2cm}

Similarly to the SEC, Eq.~(\ref{sec-eq}) implies
\begin{equation} \label{sec-ineq}
\mbox{\bf SEC} \quad \, \Longrightarrow \quad\, \dot{a} \geq a_0 \, H_0
\quad  \forall \quad a < a_0 \;.
\end{equation}  
The inequality above, along with equations~(\ref{dist-mod}) and~(\ref{dist-lumin}), furnishes
\begin{equation}  \label{SEC-bound}
\mbox{\bf SEC} \;  \Longrightarrow \; 
\mu(z) \leq 5\,\log_{10}\,
\left[\,H_0^{-1}\,(1+z)\ln (1+z)\,\right] + 25 \;.
\end{equation}

\vspace{0.2cm}

\centerline{\bf{C.  DEC}}

\vspace{0.2cm}

DEC provides an upper and a lower bound on the rate of expansion. Indeed, from~(\ref{dec-eq}) one obtains $\dot{a}\leq H_0\,a_0^3/a^2$ and $\dot{a} \geq H_0\,a$ $\forall$ $a < a_0$.
Again, these inequalities, along with~(\ref{dist-mod}) and~(\ref{dist-lumin}),
give the following upper and lower bounds for the distance modulus:
\begin{eqnarray}  \label{DEC-bound}
\mbox{\bf DEC} & \Longrightarrow & 
5\,\log_{10}\,\left[\,\frac{H_0^{-1}}{2}\,\frac{z(2+z)}{1+z} \,\right] + 25 \leq 
\mu(z) \nonumber \\
 &  & \leq 5\,\log_{10}\, \left[\,H_0^{-1}\,z(1+z)\,\right] + 25.
\end{eqnarray}
As expected [see Eqs. (\ref{wec-eq}) and (\ref{dec-eq})], the upper bound of DEC coincides with the WEC-$\mu(z)$ prediction\footnote{We note that less restrictive constraints can also be derived on the deceleration parameter, defined as $q = -a\ddot{a}/\dot{a}^2$. In this case, we find, respectively, for WEC, SEC and DEC, $q \geq -1$, $q \geq 0$, $-1 \leq q \leq 2$.}.

\section{Discussion}

Figure 1 shows the distance modulus $\mu(z)$ inequalities [Eqs. (\ref{WEC-bound}-\ref{DEC-bound})] as a function of the redshift parameter by taking $H_0^{-1} = 3000h^{-1}$ Mpc. From top to bottom the curves correspond, respectively, to the WEC, SEC and DEC predictions for $\mu(z)$. To plot these curves we have assumed two different values for $h$, i.e., $h = 0.70$ (dashed lines) and $h = 0.76$ (solid lines), which correspond to $\pm 1\sigma$ of the value provided by current CMB measurements \cite{wmap}. Note that, in the \emph{hierarchy} of the predictions for $\mu(z)$, violation of the WEC also means violation of both SEC and DEC, although the opposite does not necessarily hold true. Note also that the curves depend very weakly on the value adopted for the Hubble parameter. As an example, at $z \simeq 0.5$, the difference between the SEC-$\mu(z)$ prediction for these two values of $h$ is smaller than 1\%, which is also maintained for higher values of $z$. 

In Figure 2 we confront 
the EC predictions for $\mu(z)$ with the current SNe Ia observations. The data points appearing in the first row (Panels 2a, 2b and 2c) correspond to the so-called \emph{gold} sample of 157 events distributed over the redshift interval $0.01 \lesssim z \lesssim
1.7$ \cite{Riess2004}, whereas in the second row (Panels 2d, 2e and 2f) the data set corresponds to the first year results of the planned five years SNLS collaboration  \cite{Legacy2005}. The SNLS sample includes 71 high-$z$ SNe Ia in the redshift range $0.2 \lesssim z \lesssim 1$ and 44 low-$z$ SNe Ia ($z \leq 0.2$). In all the Panels above we have adopted $h = 0.73$, in agreement with third-year WMAP results \cite{wmap}. 

Figures 2a and 2d show the upper-bound curves $\mu(z)$ for the WEC-fulfillment. An interesting aspect of these Panels is that they seem to suggest that the WEC might have been violated by some of the nearby SNe Ia ($z \lesssim 0.2$) in both samples. As an example, let us take the cases of the SNe 1992aq, 1996ab and 1997N (from \emph{gold} sample) which are, respectively, at $z = 0.101$, $z = 0.124$ and $0.180$. While their observed distance modulus are  $\mu_{\rm{1992aq}} = 38.73 \pm 0.20$, $\mu_{\rm{1996ab}} = 39.20 \pm 0.22$ and $\mu_{\rm{1997N}} = 39.98 \pm 0.18$ \cite{Riess2004}, the upper-bound WEC predictions for the corresponding redshifts are $\mu(z = 0.101) = 38.48$, $\mu(z = 0.124) = 38.79$ and $\mu(z = 0.180) = 39.70$, respectively. A similar example of WEC violation also happens for some SNe Ia from the SLNS sample. Besides the low-$z$ cases of the SNe 1992bs and 1992bh, the high-$z$ SNLS-04D3cp at $z = 0.83$ has $\mu_{\rm{04D3cp}} = 44.414 \pm 0.347$ \cite{Legacy2005} whereas the upper-bound WEC prediction is $\mu(z = 0.83) = 43.976$, i.e., about $1.3 \sigma$ below  the central value observed by the SNLS collaboration. Note that these same considerations  are also applied to the upper-bound DEC predictions [Eq. (\ref{DEC-bound})]. Note also that the lower-bound of DEC is not violated by the current SNe Ia data (Figs. 2c and 2f).

Still on WEC and DEC violations, it is worth mentioning that the observed luminosity distance derived from SNe Ia
observations may be fitted by a dark energy component violating these energy conditions, the so-called phantom fields, first noticed by Caldwell~\cite{Caldwell2002} (see also \cite{carroll1,phantom} and references therein). However, differently from the above indications of WEC and DEC violations, most of these results are derived in a model-dependent way and under the assumption that the dark energy equation of state $\omega \equiv p/\rho$ is constant.

A similar analysis for the SEC-fulfillment is shown in Figs. 2b and 2e. For SEC, which is responsible to ensure the attractiveness of gravity and, consequently, a decelerating cosmic expansion, the plots clearly indicate a breakdown for essentially all the SNe Ia in the redshift interval covered by  \emph{gold} and SNLS samples. The interesting aspect here is that the violation of SEC happens even at very high-$z$, as indicated by the examples of the SNe 2003az (\emph{gold}) and 04D3dd (SNLS) at $z = 1.265$ and $z = 1.01$ and whose observed distance modulus are, respectively, $\mu_{\rm{2003az}} = 45.20 \pm 0.20$ and $\mu_{\rm{04D3dd}} = 44.673 \pm 0.533$. For these redshifts, the corresponding upper-bound SEC predictions are $\mu(z = 1.265) = 44.40$ and $\mu(z = 1.01) = 43.804$. The most interesting example, however, comes from the SN 2002fw (\emph{gold}) at $z = 1.3$. While its observed magnitude is $\mu_{\rm{2002fw}} = 45.27 \pm 0.19$, the upper-bound SEC prediction for the same redshift is  $\mu(z = 1.3) = 44.48$, which is about $4\sigma$ below the central value measured by the HzST collaboration.


\section{Concluding remarks}

The energy conditions  play an important role in the understanding of several properties of our Universe, including the current and past accelerating expansion phases in the context of FLRW models. They are also necessary in the formulation and proof of various singularity theorems in classical black hole thermodynamics (e.g., the proof of the second law of black hole thermodynamics requires the null energy conditions ($\rho + p \geq 0$), whereas the Hawking-Penrose singularity theorems invoke the strong energy condition, $\rho + 3p \geq 0$ and, by continuity, 
$\rho + p \geq 0$) \cite{Hawking-Ellis,Visser}.

In this paper, by using the fact that the classical energy conditions can be translated into differential constraints involving the scale factor $a$ and its derivatives [see Eqs.(\ref{wec-eq}-\ref{dec-eq})], we have derived model-independent bounds on the luminosity distance of extragalactic sources. We have confronted these energy-condition-fulfillment bounds with current SNe Ia observations from \emph{gold} and SNLS samples and shown that all the energy conditions seem to have been violated in a recent past of the cosmic evolution. The most surprising fact is that these violations may have happened even at high-$z$ [$\sim {\cal{O}}(1)$], when the Universe is expected to be dominated by usual matter fields. 

Finally, we emphasize that in agreement with other recent studies \cite{visserS}, the results reported here reinforce the idea that no possible combination of \emph{normal} matter is capable of fitting the current observational data.


\begin{acknowledgments}
The authors are grateful to N. Pires for valuable discussions. J. S. acknowledges the support of PRONEX (CNPq/FAPERN). J. S. A. and M. J. R. thank CNPq 
for the grants under which this work was carried out.

\end{acknowledgments}



\end{document}